\documentclass[12pt]{article}

\usepackage{amssymb}

\begin{document}

\begin{titlepage}

\begin{flushright}
\end{flushright}
\vskip 2.5cm

\begin{center}
{\Large \bf Lorentz Violation in Fermion-Antifermion \\
Decays of Spinless Particles}
\end{center}

\vspace{1ex}

\begin{center}
{\large Brett Altschul\footnote{{\tt baltschu@physics.sc.edu}}}

\vspace{5mm}
{\sl Department of Physics and Astronomy} \\
{\sl University of South Carolina} \\
{\sl Columbia, SC 29208} \\
\end{center}

\vspace{2.5ex}

\medskip

\centerline {\bf Abstract}

\bigskip

If Lorentz and CPT violation exist, they could affect the decays of scalar
and pseudoscalar particles. For a decay into a fermion and an antifermion
(not necessarily of the same mass), both the total decay rate and the
outgoing particle distribution may be modified, through interference
between the conventional decay mechanism and a separate Lorentz-violating
mechanism. The modifications are sensitive to forms of Lorentz violation
that are otherwise rather difficult to study, since at tree level they do
not affect particle propagation, but only interaction vertices. Using existing
experimental data on charged pion decay, it is possible to constrain three
parameters in the modified pion-muon-neutrino coupling at better than the
$10^{-9}$ level; these are the first bounds on these quantities.

\bigskip

\end{titlepage}

\newpage

\section{Introduction}

There is currently quite a bit of interest in the possibility that
Lorentz and CPT symmetries might not be exact in nature. At this point, this
is strictly a theoretical possibility. No compelling evidence has been
found that these symmetries do not hold exactly. However, if such a
discovery were made, it would obviously be of the highest importance and
would provide concrete evidence for new physics with
completely novel behavior.

Both the standard model and general relativity are invariant under Lorentz
symmetry and CPT, but deviations from these symmetries involving standard
model quanta can be described using the same kind of machinery of effective
field theory that is used in the description of the standard model itself.
The general effective field theory that can be used to describe such effects
is called the standard model extension (SME). The SME contains all
possible  translation-invariant but Lorentz-violating operators that could be
constructed out of known standard model fields. These operators break
Lorentz symmetry when they have residual tensor indices. In the Lagrangian,
these indices are contracted with tensor-valued coefficients, which describe
preferred vectors and tensors~\cite{ref-kost1,ref-kost2}.
If Lorentz symmetry is spontaneously broken, the preferred background
tensors are related to the vacuum expectation values of vector- and
tensor-valued dynamical fields.

A particular restricted version of the SME has become the standard framework
for parameterizing experimental Lorentz and CPT tests. This is the minimal SME,
which contains only local, gauge-invariant, renormalizable operators.
With these restrictions, the minimal SME has only a finite number of
undetermined parameters, and radiative corrections are calculable;  the theory
is thus suitable for making orderly comparisons of the results of different
experiments. Important tests of spatial isotropy, boost invariance, and CPT
symmetries have included studies with matter-antimatter
asymmetries for trapped charged
particles~\cite{ref-bluhm1,ref-gabirelse,ref-dehmelt1} and bound state
systems~\cite{ref-bluhm3,ref-phillips},
measurements of muon properties~\cite{ref-kost8,ref-hughes}, analyses of
the behavior of spin-polarized matter~\cite{ref-heckel3},
frequency standard comparisons~\cite{ref-berglund,ref-kost6,ref-bear,ref-wolf},
Michelson-Morley experiments with cryogenic
resonators~\cite{ref-muller3,ref-herrmann3,ref-eisele}, measurements of neutral meson
oscillations~\cite{ref-kost7,ref-hsiung,ref-abe,
ref-link,ref-aubert}, polarization measurements on the light from cosmological
sources~\cite{ref-kost21,ref-kost22},
high-energy astrophysical
tests~\cite{ref-stecker,ref-jacobson1,ref-altschul6,ref-klinkhamer2},
precision tests of gravity~\cite{ref-battat,ref-muller4}, and others.
The best current constraints based on these experiments are
collected in~\cite{ref-tables}.

The quantum electrodynamics sectors of the SME has received the most attention,
both theoretically and experimentally. In contrast, the least studied sector is
is the scalar sector. Fundamental scalars have only recently been observed
experimentally, but there are also many composite pseudoscalar particles in
the standard model. In addition to the fundamental Higgs, spin-0 mesons may
have Lorentz-violating interactions that could be described using the minimal
SME formalism. Lorentz-violating Yukawa-like interactions are a particularly rich
area; there are many possible interactions, some of which may be quite
challenging to constrain. This paper will look at the behavior of a particularly
natural process involving scalars---the production of a fermion-antifermion
pair in scalar decay. This can have particular relevance in upcoming studies
of Lorentz invariance in the Higgs sector, since the Higgs should have an
intense top quark decay mode. We shall also look specifically at the decays
of pseudoscalar mesons, and it will be possible to place strong constraints
on three pion SME coefficients which have never been studied before.

This paper is organized as follows. Section~\ref{sec-LV} introduces the
minimal SME Lagrange density that describes the scalar and spinor fields,
as well as their Lorentz-violating interactions. In
section~\ref{sec-scalardecay}, we look at several
observables---the Lorentz-violating modifications to the decay rate for a
scalar particle decaying into a fermion-antifermion pair and the
spin structure of
the decay when the daughter particles are ultrarelativistic. In
section~\ref{sec-piondecay}, we shall turn to the problem of charged pion
decay, looking at the possibility of interference between the usual
weak mechanism for the decay and a novel Lorentz-violating mechanism. It
turns out that existing experimental data can provide very strong
constraints on some of the types of Lorentz violation that might be involved.
Our conclusions and final discussion are given in section~\ref{sec-concl}.

\section{Lorentz Violation in Scalar-Spinor Interactions}
\label{sec-LV}

The coupling of bosonic and fermionic fields is more complicated in a
scalar theory than in a gauge theory. In either case, there can be various
Lorentz-violating modifications to the boson-fermion vertex. However,
in a gauge theory, the renormalizable couplings are completely determined
by the Lorentz violation in the pure fermion sector. This is a consequence of
gauge invariance, in the form of minimal coupling; the gauge field must
couple to the current, which is determined by the way the fermions propagate.
There is no such straightforward
invariance requirement in a Yukawa theory, and there exist separate
Lorentz-violation coefficients that can only be observed in processes
involving fermion-boson interaction vertices. Gauge invariance may,
however, still restrict which fields may be coupled together,
including limiting the chiralities of the spinor fields.

The most prototypical particle interaction process at high energies is
pair creation. In a theory with massive bosons, this can take the form of
a decay process. The fermion and antifermion produced in the decay may be
antiparticles---thus with equal masses---or they may be associated with
different fields, so the decay involves a net flavor change.

In our analysis of this pair production process,
we shall consider a Lorentz-violating Yukawa theory, with a minimal
SME form. With a single fermion field, the most general Lorentz-violating
Lagrange density is
\begin{equation}
\label{eq-Lf}
{\cal L}_{f}=\bar{\psi}(i\Gamma^{\mu}\partial_{\mu}-M)\psi,
\end{equation}
where the purely fermionic Lorentz violation enters through the quantities
\begin{eqnarray}
\Gamma^{\mu} & = & \gamma^{\mu}+\Gamma_{1}^{\mu}
=\gamma^{\mu}+c^{\nu\mu}\gamma_{\nu}+d^{\nu\mu}\gamma_{5}\gamma_{\nu}
+e^{\mu}+if^{\mu}\gamma_{5}+\frac{1}{2}g^{\lambda\nu\mu}
\sigma_{\lambda\nu} \\
\label{eq-LM}
M & = & m+im_{5}\gamma_{5}+M_{1}=
m+im_{5}\gamma_{5}+a^{\mu}\gamma_{\mu}+b^{\mu}\gamma_{5}\gamma_{\mu}+
\frac{1}{2}H^{\mu\nu}\sigma_{\mu\nu}.
\end{eqnarray}
These terms exhaust all the possible Lorentz structures that can be
constructed in the minimal SME. The $\Gamma$ coefficients are
dimensionless, while the $M$ coefficients have dimension $({\rm mass})^{1}$.
A number of the coefficients, such as $m_{5}$, $a$, and
$f$, can actually be eliminated from the theory, through
a redefinition of the fermion field~\cite{ref-colladay2,ref-altschul8}.
On the other hand, if more than a single fermion species is present in
the theory, there may be extra terms with these same structures that mix the
fermion fields.

However, Lorentz violation for fermions is a relatively well explored and
studied topic. Many efforts, experimental and theoretical, have been made
to constrain and understand the $M$ and $\Gamma$ coefficients. We shall only
work to first order in the Lorentz violation, so the contributions from
various SME coefficients may be determined separately and simply added together.
We shall therefore neglect the purely fermionic
coefficients and focus on the Lorentz violation
that is intrinsic to the scalar sector.

So we shall now introduce a scalar field $\phi$. With this $\phi$, the most
general boson-sector
Lagrange density with an unbroken $\phi\rightarrow-\phi$ symmetry is
\begin{equation}
\label{eq-Lb}
{\cal L}=\frac{1}{2}(\partial^{\mu}
\phi)(\partial_{\mu}\phi)+\frac{1}{2}K^{\mu\nu}(\partial_{\nu}\phi)
(\partial_{\mu}\phi)-\frac{1}{2}\mu^{2}\phi^{2}-\frac{\lambda}{4!}\phi^{4}
-\bar{\psi}_{a}G_{ab}\psi_{b}\phi.
\end{equation}
The tensor $K^{\mu\nu}=K^{\nu\mu}$ plays the same role as $c^{\nu\mu}$ in the
fermion sector, and it is the only kind of renormalizable Lorentz violation
that can be constructed with a single real scalar field.
Like $c$, $K$ can affect the kinematics for a fermion-antifermion
decay process, but $K$ does not
affect the matrix element (at tree level). Moreover, coefficients of this general
type can be interchanged through coordinate transformations (although some care
must be taken to ensure that the final observables are considered in the correct
coordinates). So the effects of $K$, like those of $c$ will be neglected here,
and we shall concentrate
instead on the forms of Lorentz violation described by $G$.

In a gauge theory, any renormalizable Lorentz violation in the boson-fermion
coupling is controlled by the same $\Gamma$ coefficients as appear in the
pure fermion sector. However, in a scalar field theory, the $G$ that appears
in the coupling is independent of the parameters appearing elsewhere in the
theory. The form of $G_{ab}$ is
\begin{equation}
G=g+ig'\gamma_{5}+G_{1}=g+ig'\gamma_{5}+I^{\mu}\gamma_{\mu}+J^{\mu}\gamma_{5}
\gamma_{\mu}+\frac{1}{2}L^{\mu\nu}\sigma_{\mu\nu}.
\end{equation}
Each of these terms carries the fermion species indices $a$ and $b$. The $a\neq b$
terms generate flavor-changing interactions. (A real $G_{ab}$ must be symmetric in
these indices in order to maintain a hermitian Lagrangian; the imaginary part of
$G_{ab}$ must likewise be antisymmetric in the flavor indices.)
The structures of the various terms contributing to $G$ mirror those seen in $M$.
The $g$ and $g'$ terms are the Lorentz-invariant scalar and pseudoscalar Yukawa
couplings, while the remaining terms are Lorentz violating. All these coefficients
are dimensionless, and the tensor term $L^{\mu\nu}$ is naturally
antisymmetric. The vector and pseudovector terms $I$ and $J$
break CPT as well as Lorentz invariance, but $L$ does not.

The existence of these terms was noted in the early descriptions of the SME,
but very little attention was paid to their rich structure until
recently~\cite{ref-ferrero3,ref-altschul30,ref-altschul35}.
However, with the recent observation of the standard model Higgs boson
at the Large Hadron Collider, the importance of fundamental scalars and 
their interactions with other species has become a central topic
in particle physics.

\section{Scalar Boson Decay Behavior}
\label{sec-scalardecay}

We shall now look at how the Lorentz-violating contributions to $G$ can
affect the decay of a spinless scalar into a fermion and an
antifermion---not necessarily with the same mass.
The simplest observable in this
case is the particle's lifetime; however, we shall see that
this turns out to be insensitive to the majority of the coefficients.
The rest frame lifetime of the scalar particle depends on $|{\cal M}|^{2}$
(where ${\cal M}$ is the invariant matrix element for the decay),
summed over the outgoing particle spins and integrated over their
directions. Of all the Lorentz-violating terms that may appear in the
scalar-spinor coupling, only the two terms $I^{0}$
and $J^{0}$ actually influence the
final rate. Purely anisotropic terms will not affect the total decay rate,
because of a cancellation between different decay channels. In the absence
of Lorentz violation, the decay of a scalar particle at rest is necessarily
isotropic. A pure anisotropy term in the decay rate, such as one
proportional to $I^{k}$, will increase the likelihood of certain decay
channels---through a dependence on, for example, $I^{k}p_{k}$ (where
$\vec{p}$ is one of the daughter particle momenta). However,
decay channels with different momenta will have their importances
correspondingly curtailed, leaving the net decay rate unchanged.

The decay rate depends on the quantity
\begin{equation}
{\cal M}=\bar{u}_{1}(p_{1})G_{12}v_{2}(p_{2}).
\end{equation}
This represents the amplitude for the scalar to decay into a
fermion with momentum $p_{1}$ and an antifermion with momentum
$p_{2}$.
Squaring this gives
\begin{equation}
|{\cal M}|^{2}=\bar{u}_{1}(p_{1})G_{12}v_{2}(p_{2})\bar{v}_{2}(p_{2})
\bar{G}_{12}u_{1}(p_{1}),
\end{equation}
where $\bar{G}_{ab}=\gamma_{0}G^{\dag}_{ab}\gamma_{0}$ is the usual
Dirac conjugation of $G_{ab}$. The Dirac conjugate simply replaces the
$g$, $g'$, $I$, $J$, and $L$ coefficients with their complex conjugates;
this only affects the coefficients that are off diagonal in flavor
space ($a\neq b$).

Since changes to the fermion sector have been neglected,
the sums over the outgoing spins can be evaluated using standard fermion
closure relations. This gives (dropping the flavor indices)
\begin{eqnarray}
\sum_{s_{1},s_{2}}|{\cal M}|^{2} & = & {\rm tr}\left\{(\!\not\!p_{1}+m_{1})
\left(g+ig'\gamma_{5}+I^{\mu}\gamma_{\mu}+J^{\mu}\gamma_{5}\gamma_{\mu}+
\frac{1}{2}L^{\mu\nu}\sigma_{\mu\nu}\right)\right. \nonumber\\
& & \left.(\!\not\!p_{2}-m_{2})
\left(g^{*}+ig'^{*}\gamma_{5}+I^{\mu*}\gamma_{\mu}+J^{\mu*}\gamma_{5}
\gamma_{\mu}+\frac{1}{2}L^{\mu\nu*}\sigma_{\mu\nu}\right)
\right\}.
\end{eqnarray}
Expanded to leading order in the $G_{1}$ coefficients, this is
\begin{eqnarray}
\sum_{s_{1},s_{2}}|{\cal M}|^{2} & = & {\rm tr}\left\{(\!\not\!p_{1}+m_{1})
(g+ig'\gamma_{5})(\!\not\!p_{2}-m_{2})(g^{*}+ig'^{*}\gamma_{5})\right\}
\nonumber\\
& & +{\rm tr}\Bigg\{(\!\not\!p_{1}+m_{1})\left(I^{\mu}\gamma_{\mu}+
J^{\mu}\gamma_{5}\gamma_{\mu}+\frac{1}{2}L^{\mu\nu}\sigma_{\mu\nu}\right)
(\!\not\!p_{2}-m_{2})(g^{*}+ig'^{*}\gamma_{5}) \nonumber\\
\label{eq-Mlinear}
& & +\left(G\leftrightarrow G^{*},p_{1}\leftrightarrow p_{2},
m_{1}\leftrightarrow -m_{2}\right)\Bigg\}.
\end{eqnarray}
The designation $G\leftrightarrow G^{*}$ indicates that the last
term in the sum differs from the previous one by the replacement of the 
$g$, $g'$, $I$, $J$, and $L$ coefficients by their complex
conjugates (as well as the interchanges of momenta and masses
also indicated).
Evaluating the traces yields
\begin{eqnarray}
\sum_{s_{1},s_{2}}|{\cal M}|^{2} & = & 4(|g|^{2}+|g'|^{2})p_{1}\cdot p_{2}
-4(|g|^{2}-|g'|^{2})m_{1}m_{2} \nonumber\\
& & -4(g^{*}I^{\mu}+gI^{\mu*})(m_{2}p_{1\mu}-m_{1}p_{2\mu})
+4i(g'^{*}J^{\mu}-g'J^{\mu*})(m_{2}p_{1\mu}-m_{1}p_{2\mu}) \nonumber\\
\label{eq-Msquared}
& & +4i(g^{*}L^{\mu\nu}-gL^{\mu\nu*}
+g'^{*}\tilde{L}^{\mu\nu}-g'\tilde{L}^{\mu\nu*})
p_{1\mu}p_{2\nu},
\end{eqnarray}
where $\tilde{L}^{\mu\nu}=\frac{1}{2}\epsilon^{\mu\nu\alpha\beta}
L_{\alpha\beta}$ is the dual of $L$.

This expression has a number of noteworthy features. In the rest
frame of the decaying scalar, $\vec{p}_{1}=-\vec{p}_{2}$, and when
$|{\cal M}|^{2}$ is integrated over all decay channels, all the
$I$ and $J$ contributions except those related to $I^{0}$ and
$J^{0}$ cancel out. Moreover, even these
isotropic contributions are nonzero
only for flavor-changing decay modes, for which the daughter masses and the
time components of $p_{1}$ and $p_{2}$ are unequal.

The energy component of $p_{1}$
is $E_{1}=\frac{\mu}{2}+\frac{m_{1}^{2}}{2\mu}-\frac{m_{2}^{2}}{2\mu}$,
and the corresponding momentum is
$|\vec{p}_{1}|=\frac{1}{2\mu}\sqrt{(\mu^{2}-m_{1}^{2}-m^{2}_{2})^{2}
-4m_{1}^{2}m_{2}^{2}}$. Consequently, the time component of
$m_{2}p_{1\mu}-m_{1}p_{2\mu}$
\begin{equation}
\label{eq-0component}
m_{2}E_{1}-m_{1}E_{2}=\left[\frac{\mu}{2}-\frac{(m_{1}+m_{2})^{2}}{2\mu}
\right](m_{2}-m_{1}).
\end{equation}
The modified decay rate is affected by (\ref{eq-0component}), times
the real part of $g^{*}I^{0}$ plus the imaginary part of
$g'^{*}J^{0}$. These contributions are not invariant under Lorentz
boosts, so they could be measured by observing the decay rates of
particles with different velocities. This strategy will be discussed in
more detail in section~\ref{sec-piondecay}.

Since $L$ has no isotropic part, we should expect it to make no
contribution at all to the decay rate. This is indeed the case, but the
structure of the $L$ terms merits some additional comment. From the
structure of the $L$ term in (\ref{eq-Msquared}), it is evident that
only the $L^{0j}$ and $\tilde{L}^{0j}$ terms contribute in the scalar
rest frame, since $L_{jk}$ is antisymmetric
and the spatial momenta $\vec{p}_{1}$ and $\vec{p}_{2}$
are collinear. However, since these six components of $L$ and
$\tilde{L}$ actually contain all six $L$ coefficients, the
differential decay rate (as a function of the direction) can
still depend on all the $L$ parameters. The key momentum
quantity is
\begin{equation}
E_{1}p_{2j}-E_{2}p_{1j}=-\mu p_{1j},
\end{equation}
which gives a dipolar assymmetry in the decay but clearly does not
contribute to the total rate when $\vec{p}_{1}$ is integrated over all
angles.

The overall anisotropy of the differential decay rate $d\Gamma/d\Omega$
as a function of the fermion direction $\hat{p}_{1}$ is
\begin{equation}
\frac{d\Gamma/d\Omega}{(d\Gamma/d\Omega)_{0}}=1+
\frac{\sqrt{(\mu^{2}-m_{1}^{2}-m^{2}_{2})^{2}
-4m_{1}^{2}m_{2}^{2}}}
{|g|^{2}[\mu^{2}-(m_{1}+m_{2})^{2}]+
|g'|^{2}[\mu^{2}-(m_{1}-m_{2})^{2}]}
W_{j}\hat{p}_{1j},
\end{equation}
where
\begin{equation}
W_{j}=2\frac{m_{1}+m_{2}}{\mu}(\Re\{g^{*}I_{j}\}+\Im\{g'^{*}J_{j}\})-
\Im\{gL_{0j}+g'\tilde{L}_{0j}\}.
\end{equation}
The vector $\vec{W}$ governs the anisotropy in a particular
rest frame. While it is not possible, using only this decay
process, to disentangle the contributions from $I$ from those
of $J$ (provided both of the conventional couplings $g$ and $g'$
are nonzero), it is
possible to separate the $L$ terms, because the $L_{0j}$ and
$\tilde{L}_{0j}$ behave differently under Lorentz boosts
than do the $I_{j}$ and $J_{j}$.

Finally, we note that for a decay into a fermion and its 
equal-mass antiparticle, $G_{1}$
makes zero total contribution to the decay rate. Because the
coefficients must be real, the $J$ and $L$ terms vanish identically.
Because of the equality of the fermion and antifermion masses, the
$I^{0}$ term vanishes as well. The only term in (\ref{eq-Msquared})
that can be nonzero is $16gmI^{j}p_{1j}$. Therefore the only effect
on this kind of decay is that the fermion is preferentially emitted
along a direction parallel to $\vec{I}$, so the antifermion is
preferentially emitted in the antiparallel direction.

So far, we have worked only with the spin-summed decay rate,
ignoring the polarizations of the decay products.
It is also possible to perform a full analysis, including the spin
dependence of the decay amplitudes, at ultrarelativistic energies
(large enough that $m_{1}$ and $m_{2}$ may be neglected).
This is accomplished by inserting the ultrarelativistic spin
projector $\frac{1}{2}(1\pm\gamma_{5})$ into the expression 
for the matrix element. In this case, the
amplitude squared is
\begin{eqnarray}
|{\cal M}|^{2} & = & {\rm tr}\left\{\frac{1}{2}(1+s_{1}\gamma_{5})
\!\not\!p_{1}
\left(g+ig'\gamma_{5}+I^{\mu}\gamma_{\mu}+J^{\mu}\gamma_{5}\gamma_{\mu}+
\frac{1}{2}L^{\mu\nu}\sigma_{\mu\nu}\right)\right. \nonumber\\
& & \left.\times\frac{1}{2}(1-s_{2}\gamma_{5})\!\not\!p_{2}
\left(g^{*}+ig'^{*}\gamma_{5}+I^{\mu*}\gamma_{\mu}+J^{\mu*}
\gamma_{5}\gamma_{\mu}+
\frac{1}{2}L^{\mu\nu*}\sigma_{\mu\nu}\right)\right\}.
\end{eqnarray}
In this expression, $s_{1}$ and $s_{2}$ represent the fermion and
antifermion helicities, respectively.
It is quickly evident that the only Lorentz-violating
vertex terms that can contribute are those involving the $L$
coefficients. When the masses are neglected, any term involving a
single factor of $I$ or $J$ must necessarily have an odd number of
$\gamma$-matrices inside the trace.

Taking advantage of this fact, and the fact that
$g+ig'\gamma_{5}+\frac{1}{2}L^{\mu\nu}\sigma_{\mu\nu}$ commutes
with $\gamma_{5}$, the $s_{1}$ and $s_{2}$
projectors may be combined into single expressions of the form
$\frac{1}{4}(1+s_{1}s_{2})(1\pm s_{1}\gamma_{5})$. Then the expression
for $|{\cal M}|^{2}$ is
\begin{eqnarray}
|{\cal M}|^{2} & = & \frac{1}{4}{\rm tr}\left\{\!\not\!p_{1}
(g+ig'\gamma_{5})\!\not\!p_{2}(g^{*}+ig'^{*}\gamma_{5})
(1+s_{1}s_{2})(1+s_{1}\gamma_{5})\right\}
\label{eq-Mspin}
\\
& & +\frac{1}{8}{\rm tr}\{\!\not\!p_{1}L^{\mu\nu}
\sigma_{\mu\nu}\!\not\!p_{2}(g^{*}+ig'^{*}\gamma_{5})
(1+s_{1}s_{2})(1+s_{1}\gamma_{5}) \nonumber\\
& & +\!\not\!p_{2}L^{\mu\nu*}
\sigma_{\mu\nu}\!\not\!p_{1}(g+ig'\gamma_{5})
(1+s_{1}s_{2})(1-s_{1}\gamma_{5})\}.
\end{eqnarray}
The Lorentz-invariant term in (\ref{eq-Mspin}) is straightforward.
There is an overall factor of $(1+s_{1}s_{2})$, which ensures
that the outgoing particles have the same helicity, as they should
for a state with vanishing total angular momentum. (Although
isotropy is broken by the SME interactions, transitions that
actually violate angular momentum conservation must have rates that
are quadratic in the SME coefficients.)

The rate for the
Lorentz-invariant process is just set by the magnitude of the
coupling
$|g+is_{1}g'|$ for the outgoing spin state being considered.
The $L$-dependent term has exactly the same structure as the $L$ term
in (\ref{eq-Mlinear}), with the replacements $g\rightarrow(g-is_{1}g')$
and $g'\rightarrow(is_{1}g+g')$. The trace manipulations can be
carried over directly, so the final result for the matrix element
squared is
\begin{eqnarray}
|{\cal M}|^{2} & = & (1+s_{1}s_{2})\left|g+is_{1}g'\right|^{2}
p_{1}\cdot p_{2}
\nonumber\\
& & +i(1+s_{1}s_{2})[(g^{*}+is_{1}g'^{*})L^{\mu\nu}-
(g-is_{1}g')L^{\mu\nu*} \nonumber\\
\label{eq-M2spin}
& & +(g'^{*}-is_{1}g^{*})\tilde{L}^{\mu\nu}-
(g'+is_{1}g)\tilde{L}^{\mu\nu*}]
p_{1\mu}p_{2\nu}.
\end{eqnarray}

This expression does not depend on $I$ or $J$, only on $L$.
The lack of
dependence on these parameters is not surprising. The expression for
$|{\cal M}|^{2}$ is ${\cal O}(p_{1}p_{2})$ because of the neglect of the
mass terms. There is clearly no structure that is linear in both outgoing
momenta but which has only a single Lorentz index to be contracted with
$I^{\mu}$ or $J^{\mu}$. This is the ultimate reason why the $I$ and
$J$ terms in the trace all involved odd number of Dirac matrices.

%

\section{Modifications to Pion Decay}
\label{sec-piondecay}

Thus far, we have considered interference between conventional and
Lorentz-violating decay mechanisms for a fundamental scalar field. In
these scenarios, the Lorentz-symmetric interactions always
had a Yukawa-like structure. This kind of analysis
is appropriate for studies of Higgs interactions, although Higgs decays
are complicated by the additional electroweak couplings of the Higgs
sector and by the fact that scalar Higgs decays frequently involve quarks,
which are affected by the strong interactions.

In this section, we shall look at how the Lorentz-violating $G_{1}$ couplings
can modify a different kind of standard model decay process---the leptonic
decay of a meson. For definiteness, we shall take the decaying particle to
be a pion, although the theoretical analysis could obviously be applied to
the weak decays of heavier mesons as well. However, there is actually fairly
good data on the boost dependence of the relevant pion decay rate, and this
data may be used to constrain the $G_{1}$ coefficients for the pion-muon-neutrino
vertex.

The pion is obviously not a fundamental field, and its
standard model interaction with the
leptons is ultimately mediated by a $W$ boson. However, there is a
well-known effective vertex that describes this process, since the virtual
$W$ involved is very far off shell. A mechanism with a pion-muon-neutrino
$G_{1}$ vertex could produce the same outgoing states, and the two
amplitudes for the conventional and Lorentz-violating reaction paths would
interfere. In this case, the $G_{1}$
involved should also be seen as representing
an effective vertex. The $I$, $J$, and $L$ involved would be linear
combinations of other tensor coefficients
associated with the fundamental standard model fields.

The effective interaction Lagrange
density for the weak decay of a pion is
\begin{equation}
\label{eq-weakvert}
{\cal L}_{int}=-iG_{F}\cos\theta_{C}
f_{\pi}\frac{\partial\phi}{\partial x_{\mu}}
\bar{\psi}_{1}\gamma_{\mu}(1-\gamma_{5})\psi_{\nu}
\end{equation}
(plus an obvious hermitian conjugate). The coupling involves the
Fermi constant $G_{F}$, the Cabibbo angle $\theta_{C}$, and the pion
decay constant $f_{\pi}$.
The charged pion ($\phi$), muon ($\psi_{1}$), and massless neutrino
($\psi_{2}$)
fields all have the Lorentz-invariant free propagation Lagrangians used
previously, and we shall consider the same kind of Lorentz-violating
$G_{1}$ interaction shown in (\ref{eq-Lb}).
Because the weak interaction only couples to the left chiral current,
Lorentz invariance requires that the standard model coupling between
pion, muon, and neutrino have an extra momentum factor to be contracted
with the current; this gives the effective interaction dimension
$($mass$)^{5}$. However, a Lorentz-violating effective vertex is not so
constrained and may be of a lower-dimensional, renormalizable type.

The matrix element squared for the charged pion decay, using both
the effective vertex for the $W$ exchange and the novel Lorentz-violating
vertex, is (using $\kappa=G_{F}\cos\theta_{C}f_{\pi}$)
\begin{eqnarray}
|{\cal M}|^{2} & = & \left|\bar{u}(p_{1})\left[\kappa
(\!\not\!p_{1}+\!\not\!p_{2})(1-\gamma_{5})+I^{\mu}\gamma_{\mu}+
J^{\mu}\gamma_{5}\gamma_{\mu}+\frac{1}{2}L^{\mu\nu}\sigma_{\mu\nu}
\right]v(p_{2})\right|^{2} \\
\sum_{s_{1},s_{2}}|{\cal M}|^{2} & = & \kappa^{2}\,{\rm tr}\left\{
(\!\not\!p_{1}+m_{1})(\!\not\!p_{1}+\!\not\!p_{2})(1-\gamma_{5})
\!\not\!p_{2}(\!\not\!p_{1}+\!\not\!p_{2})(1-\gamma_{5})\right\}
\nonumber \\
& & +\kappa\,{\rm tr}\left\{(\!\not\!p_{1}+m_{1})(\!\not\!p_{1}+
\!\not\!p_{2})(1-\gamma_{5})\!\not\!p_{2}\left(I^{\mu*}\gamma_{\mu}
+J^{\mu*}\gamma_{5}\gamma_{\mu}+\frac{1}{2}L^{\mu\nu*}\sigma_{\mu\nu}
\right)\right. \nonumber\\
& & \left.+(\!\not\!p_{1}+m_{1})\left(I^{\mu}\gamma_{\mu}
+J^{\mu}\gamma_{5}\gamma_{\mu}+\frac{1}{2}L^{\mu\nu}\sigma_{\mu\nu}
\right)\!\not\!p_{2}(\!\not\!p_{1}+\!\not\!p_{2})(1-\gamma_{5})
\right\}.
\end{eqnarray}
The required trace manipulations are very similar to those in the
previous calculations. The result is (using the on-shell conditions
$p_{1}^{2}=m_{1}^{2}=m_{\mu}^{2}$ and $p_{2}^{2}=0$)
\begin{eqnarray}
\sum_{s_{1},s_{2}}|{\cal M}|^{2}
& = & 4\kappa^{2}m_{1}^{2}(p_{1}\cdot p_{2})
+4\kappa m_{1}^{2}p_{2}\cdot(I+I^{*}+J+J^{*}) \nonumber\\
& & +4i\kappa m_{1}
(L^{\mu\nu}-L^{\mu\nu*})p_{1\mu}p_{2\nu}+4\kappa m_{1}
(\tilde{L}^{\mu\nu}+\tilde{L}^{\mu\nu*})p_{1\mu}p_{2\nu}.
\end{eqnarray}
Note that, of the $I$ and $J$ parameters, only the left-chiral
combination $I_{L}^{\mu}
=I^{\mu}+J^{\mu}$ enters, since the weak process only involves
the left-chiral fermion fields.

The decay rate depends only on the average of $|{\cal M}|^{2}$ over
all outgoing particle directions. Only those terms which are isotropic
in the pion rest frame contribute to this average, so
\begin{equation}
|{\cal M}|^{2}_{ave}=4\kappa^{2}m_{1}^{2}(p_{1}\cdot p_{2})+
4\kappa m_{1}^{2}E_{2}(I_{L}^{0}+I_{L}^{0*}).
\end{equation}
The kinematics of the decay have $E_{1}=[(1+m_{\mu}^{2}/m_{\pi}^{2})/2]
m_{\pi}$ and $|\vec{p}_{1}|=|\vec{p}_{2}|=E_{2}=
[(1-m_{\mu}^{2}/m_{\pi}^{2})/2]m_{\pi}$, so that
$p_{1}\cdot p_{2}=E_{2}m_{\pi}$. This simplifies the average
$|{\cal M}|^{2}$---to which the decay rate $\Gamma$ is directly
proportional---so that
\begin{equation}
\Gamma=\Gamma_{0}\left(1+\frac{I_{L}^{0}+I_{L}^{0*}}{\kappa m_{\pi}}\right)
=\Gamma_{0}\left(1+\frac{I_{L}^{0}+I_{L}^{0*}}{G_{F}\cos\theta_{C}
f_{\pi}m_{\pi}}\right).
\end{equation}
Numerically, this is $\Gamma=\Gamma_{0}[1+(7.3\times 10^{6})
(I_{L}^{0}+I_{L}^{0*})]$, indicating that the Lorentz-violating
effect is subject to a large enhancement factor.

The enhancement arises from the interference between decay
amplitudes produced by two different dynamical mechanisms.
In previously studied scenarios involving Lorentz-violating changes to this
decay~\cite{ref-nielsen,ref-altschul32,ref-altschul33,ref-altschul34}
(as well as most other studies of Lorentz violation in
elementary particle processes),
the interaction responsible for the process was
the conventional, minimally coupled standard
model gauge interaction. The Lorentz-violating
modifications in previous analyses were essentially changes to the
propagation structure for the various fields involved in the decay. These
could include Lorentz violation for the parent meson, the daughter
leptons, and even the $W$ boson that mediates the weak decay. Such
modifications are not just kinematic in nature; they can affect the
invariant matrix element ${\cal M}$ for the weak decay process, through
changes to the propagators for the internal particles 
and also through the changes to the vector boson vertices that are
demanded by gauge invariance.

However, the $G_{1}$ interaction is something else, entirely new, with
a strength that is not tied in any way to the strength of the
conventional interaction
The normal pion decay is rather slow, because the weak interaction
is not particularly strong at low energies. So a small $G_{1}$ might
still produce a sizable fractional change in $\Gamma$, because
$\Gamma_{0}$ is simply quite small to begin with. Even a modestly accurate
measurement of the fractional change in $\Gamma$ can thus produce a
very strong constraint on the $I_{L}$ parameters; the weakness of the
standard model process enhances the sensitivity to a new,
Lorentz-violating mechanism.

Although the decay rate depends
solely on the component of $I_{L}$ that is isotropic
in the pion rest frame, this component $I_{L}^{0}$ does depend strongly on the
speed and direction of the pion's motion. Experimental bounds on SME parameters
are conventionally given in a particular system of Sun-centered celestial
equatorial coordinates $(X,Y,Z,T)$~\cite{ref-bluhm4}. The origin of the
coordinate system is positioned at the center of the Sun. The coordinate
$Z$-axis points parallel to the Earth's rotation axis; the $X$-axis
points in the direction of the vernal equinox point on
the sky; and naturally the
$Y$-axis is  determined by the right hand rule. The conventional time coordinate
is denoted by $T$; however, it is frequently advantageous to use a translated
local time coordinate $T_{\oplus}$, selected so that
at $T_{\oplus}=0$, the $y$-direction in the laboratory coincides with the
$Y$-direction in the reference coordinates.

The coefficient $I_{L}^{0}$ in the pion rest frame is related to the
coefficients in the Sun-centered frame by a boost,
\begin{equation}
I_{L}^{0}=\gamma_{\pi}\left[(I_{L})_{T}+\left(\hat{v}_{\pi}
\right)_{J}(I_{L})_{J}\right],
\end{equation}
where $\gamma_{\pi}\gg1$ and $\hat{v}_{\pi}$ are the Lorentz factor and
direction for the pion's motion, respectively.
A large value of $\gamma_{\pi}$ provides
further enhancement of the sensitivity of the experiment. The reason for
this enhancement is that a test of Lorentz symmetry must ultimately involve
a comparison of equivalent observables in differently oriented or boosted
frames. A large pion speed means that, as the beam direction changes, a
collection of very different rest frames is being sampled.

Notably, there are 
already experimental constraints on how the pion lifetime might depend on
the boost of the decaying particle's rest frame. The underlying
experimental data was collected by the MINOS experiment
and analyzed as part of a
search for possible Lorentz-violating neutrino
oscillations~\cite{ref-adamson1}. There was
found to be no evidence of any sidereal oscillations in the number of
charged current events in the MINOS near detector. Such oscillations
would be a signature of Lorentz violation in the neutrino oscillation
probability, but it has been noted~\cite{ref-altschul33} that
they would also be the signature of Lorentz violation
in the pion decay rate. If its lifetime depended on the direction
in which a pion is moving, then the overall intensity of the NuMI beam
would depend on the orientation of the beam apparatus and thus the time
of day.

Note that only the net decay rate is observable via this method. Lorentz
violation would generally produce a decay with an anisotropic distribution
of decay products in the pion rest frame. However, this anisotropy is
effectively washed out by the fact that essentially all the daughter
particles are beamed into a narrow pencil of angles along the NuMI beam
direction. Moreover, the dependence of the overall NuMI neutrino beam
strength on the pion lifetime is a bit more complicated than one might
initially anticipate~\cite{ref-altschul33}.
The length of the pion decay pipe is an important parameter.
If the pipe were very long, essentially all the pions in the initial pion
beam have time to decay; this would make the neutrino beam intensity
almost independent of the pion lifetime. On the other hand, if the pipe
were very short, any change in the pion decay rate would be precisely
mirrored with the same fractional change
in the neutrino beam strength, but the beam so
produced would be very weak and the experimental statistics correspondingly
poor. For MINOS, where the pipe length was comparable to the mean
decay length for the pions involved, a change in the observed beam
intensity $I$ is related to a change in the pion decay rate $\Gamma$ as
\begin{equation}
\frac{\Delta I}{I_{0}}=
\frac{\Gamma_{0}D/\gamma_{\pi}}{e^{\Gamma_{0}D/\gamma_{\pi}}-1}
\left(\frac{\Delta \Gamma}{\Gamma_{0}}\right)=0.31
\left(\frac{\Delta \Gamma}{\Gamma_{0}}\right),
\end{equation}
where $D$ is the pipe length, and the numerical
value was calculated for a typical 6.0 GeV pion.

The geometry of the NuMI beam also determines how the pion boost varies with
time. The whole apparatus rotates with the planet, which will produce
sidereal oscillations in the key parameter $I_{L}^{0}$. The pion decays occur 
at colatitude $\chi=42.18^{\circ}$, and the spherical coordinates $(\theta,\phi)$
describe the angle between the beam orientation and the local zenith direction
($\theta=93.27^{\circ}$) and the azimuthal angle in the plane of the Earth's surface,
measured starting eastward from south ($\phi=203.91^{\circ}$).
Consequently, the beam direction
at a local time $T_{\oplus}=0$ is~\cite{ref-mewes2}
\begin{eqnarray}
\hat{v}_{\pi} & = & N_{1}\hat{X}+N_{2}\hat{Y}+N_{3}\hat{Z} \\
& = & (\cos\chi\sin\theta\cos\phi+\sin\chi\cos\theta)\hat{X}
+(\sin\theta\sin\phi)\hat{Y} \nonumber\\
& & +(-\sin\chi\sin\theta\cos\phi+\cos\chi\cos\theta)
\hat{Z} \\
& = & -0.715\hat{X}-0.405\hat{Y}+0.571\hat{Z}.
\end{eqnarray}
As the Earth rotates with sidereal frequency $\omega_{\oplus}$, the crucial parameter
$I_{L}^{0}$ varies according to
\begin{eqnarray}
I_{L}^{0} & = & \gamma_{\pi}\left\{\left[(I_{L})_{T}+N_{3}(I_{L})_{Z}\right]
+\left[N_{1}(I_{L})_{X}+N_{2}(I_{L})_{Y}\right]\cos(\omega_{\oplus}T_{\oplus})
\right.
\nonumber\\
\label{eq-I0osc}
& & \left.
+\left[-N_{2}(I_{L})_{X}+N_{1}(I_{L})_{Y}\right]\sin(\omega_{\oplus}T_{\oplus})
\right\}.
\end{eqnarray}
Because the rotation of the Earth does not change the $Z$-component of $\hat{v}_{\pi}$,
$(I_{L})_{Z}$ does not affect the sidereal oscillation amplitude.

Ultimately, the amplitudes of any oscillations in the beam strength are given by
the real parts of the sine and cosine coefficients
inside the square brackets in (\ref{eq-I0osc}), times a sensitivity factor
\begin{equation}
{\cal S}=2\gamma_{\pi}\left(\frac{1}{G_{F}\cos\theta_{C}f_{\pi}m_{\pi}}\right)
\left(\frac{\Gamma_{0}D/\gamma_{\pi}}{e^{\Gamma_{0}D/\gamma_{\pi}}-1}\right)
=1.9\times10^{8}.
\end{equation}
The MINOS near detector charge current event rate showed no evidence of oscillations
reaching the $3\sigma$ level of significance (a level that could be seen as
indicative of an actual effect). In fact, the levels of signal present
in the data were well below the $3\sigma$ threshold. However, even with no
positive evidence for a signal, it was still possible to calculate the
the Fast Fourier Transform
(FFT) power present in the $\sin(\omega_{\oplus}T_{\oplus})$
and $\cos(\omega_{\oplus}T_{\oplus})$ modes (as well as in higher harmonics that
are not relevant for the present analysis)~\cite{ref-adamson1}. The statistical
noise that was present in these quadrature modes was characteristic of
a fractional  $1\sigma$ dispersion of $1.8\times10^{-2}$ in the FFT power.

Since the MINOS group's analysis looked only at one source of Lorentz violation
at a time, we shall take the same approach here. In order that an actual
Lorentz-violating $\Re\{(I_{L})_{J}\}$ (for $J=X$ or $Y$) should go undetected
amidst the noise in the data, the quantity $|{\cal S}N_{1}\Re\{(I_{L})_{J}\}|$
would need to be less than the $3\sigma$ noise level of $5.4\times10^{-2}$.
This gives us our final bounds on the coefficients involved,
\begin{equation}
\label{eq-ILJbound}
|\Re\{(I_{L}^{\pi\mu\nu})_{J}\}|<4\times10^{-10}
\end{equation}
for $J=X,Y$. The superscripts on $I_{L}$ denote that these are coefficients for
the pion-muon-neutrino vertex. The results (\ref{eq-ILJbound}) represent
two strong constraints on coefficients that
have never previously been bounded.

The pion decay rate is also sensitive to violations purely of boost
invariance, through $(I_{L})_{T}$. However, this requires comparisons of
pion lifetimes for particles with different boost factors, and most precise
measurements of the lifetime use stopped pions. Assuming that only
$\Re\{(I_{L})_{T}\}$ is nonzero and using the same two data
points~\cite{ref-nordberg,ref-ayers}
(at $\gamma_{\pi}=1$ and $\gamma_{\pi}=2.44$)
discussed in~\cite{ref-nielsen}, the differences in rest frame lifetimes between
stopped pions and pions in motion provide the result
\begin{equation}
\label{eq-ILTbound}
|\Re\{(I_{L}^{\pi\mu\nu})_{T}\}|=(7\pm4)\times10^{-10}.
\end{equation}
So the isotropic boost invariance violation term in $I_{L}$ and two of the three
anisotropic terms are constrained at similar levels.

\section{Conclusion}
\label{sec-concl}

The constraints (\ref{eq-ILJbound}--\ref{eq-ILTbound})
are the first bounds published on any of the
$G_{1}$ coefficients of this type. These Yukawa-like Lorentz-violating parameters
are quite numerous in the SME, but they have not been extensively studied.
It has been established that the $I$, $J$, and $L$ play roles in the scalar
potentials between fermionic particles and in
the renormalization group flow of other SME coefficients. However, neither of
these observations has been developed in sufficient detail to make it possible to
place practical experimental bounds on the $G_{1}$ parameters.
Moreover, such bounds would generally be expected to be
rather poor, because of their inherently indirect nature. On the other hand, the
constraints given here are quite strong. While the MINOS detector data was not
particularly sensitive to sidereal oscillations in the neutrino beam strength,
the sensitivity to the relevant $I_{L}$ coefficients was enhanced by two
important factors. The relativistic motion of the decaying pions added a
factor of $\gamma_{\pi}$, and the weakness of the standard model process meant
that a small SME contribution could still make a sizable correction to the
net decay rate. Measurements of lepton universality, comparing the branching
ratios for meson decays into first- versus second-generation leptons, could
also provide sensitivity to differences among the $I_{L}$ parameters.

The pion, of course, is not a fundamental particle, but a composite made from
quarks and gluons. In this paper, we have looked at the two-particle decays
both of fundamental scalars, with only renormalizable interactions, and
composites, with weak interactions of the type (\ref{eq-weakvert}). While the
standard model certainly contains fundamental scalars---which are only
beginning to be studied directly---there also exists a
profusion of spin-0 species at the hadronic scale.
The fields representing these composite
particles have Lorentz-violating operators associated with them in effective
field theory. The operators that were constrained in section~\ref{sec-piondecay}
were of this type. However, while constraints on these and other
effective parameters for the pion sector~\cite{ref-altschul16} are certainly
interesting, it would desirable to understand how the coefficients for
particles like pions, protons, and neutrons are
related to the coefficients for the underlying quark and gluon fields.
In fact, understanding the effective coefficients for hadrons is one of the most
important outstanding theoretical problems in the SME.

Understanding Yukawa-like Lorentz violating interactions for the fundamental
spin-0 fields
will also be important in the coming years, as the detailed behavior of the
Higgs particle is mapped out. As it becomes possible to test the Lorentz
and CPT invariances of the standard model's scalar sector, the $G_{1}$
coefficients should be an important part of that analysis. For the standard
model Higgs, the only $G_{1}$ coefficients that can appear are the $L$
coefficients, because the Higgs field is part of a doublet representation of
the $SU(2)_{L}$ gauge field and so can only be coupled to another doublet.
While searching for Lorentz violation in the Higgs sector may seem
like a difficult undertaking, experimental studies of Lorentz invariance for
the similarly heavy and short-lived top quark have already been
conducted~\cite{ref-abazov}.

This paper has laid groundwork for future studies of Lorentz
violation involving scalars, by looking at
two-body decays mediated by the Lorentz-violating interactions. The decay
rates typically depend on the directions in which both the parent and
daughter particles are moving, and both CPT-even and CPT-violating signatures
have been identified, associated with different coefficients.

\end{document}